\documentclass[twocolumn,preprintnumbers,amsmath,amssymb,superscriptaddress]{revtex4-2}
\usepackage{graphicx}
\usepackage{dcolumn}
\usepackage{bm}
\usepackage{color}
\usepackage{ulem}
\usepackage{hyperref}

\font\scripti=cmmi7
\font\scriptscripti=cmmi5
\def\sib#1{\setbox0 = \hbox{\scripti #1}
  \kern-.02em\copy0\kern-\wd0
  \kern.04em\box0} 
\def\ssib#1{\setbox0 = \hbox{\scriptscripti #1}
  \kern-.02em\copy0\kern-\wd0
  \kern.04em\box0} 
\font\tenib=cmmib10 
\skewchar\tenib='177 \skewchar\tenib='177 \skewchar\tenib='177
\def\pbold#1{\setbox0 = \hbox{$ #1 $}
  \kern-.022em\copy0\kern-\wd0
  \kern.011em\copy0\kern-\wd0
  \kern.011em\copy0\kern-\wd0
  \kern.011em\copy0\kern-\wd0
  \kern.011em\box0} 

\usepackage{graphicx}
\usepackage{dcolumn}
\usepackage{bm}
\usepackage{mathrsfs}

\def\lesssim{\ \raise.3ex\hbox{$<$}\kern-0.8em\lower.7ex\hbox{$\sim$}\ }
\def\gesim{\ \raise.3ex\hbox{$>$}\kern-0.8em\lower.7ex\hbox{$\sim$}\ }

\begin{document}
\preprint{RIKEN-iTHEMS-Report-22}
\title{Non-Hermitian topological Fermi superfluid near the {\it p}-wave unitary limit}
\author{Hiroyuki Tajima}
\affiliation{Department of Physics, Graduate School of Science, The University of Tokyo, Tokyo 113-0033, Japan}
\author{Yuta Sekino}
\affiliation{RIKEN Cluster for Pioneering Research (CPR), Astrophysical Big Bang Laboratory (ABBL), Wako, Saitama, 351-0198 Japan}
\affiliation{Interdisciplinary Theoretical and Mathematical Sciences Program (iTHEMS), RIKEN, Wako, Saitama 351-0198, Japan}
\author{Daisuke Inotani}
\affiliation{Departments of Physics, Keio University, 4-1-1 Hiyoshi, Kanagawa 223-8521, Japan}
\author{Akira Dohi}
\affiliation{Department of Physics, Hiroshima University, Higashi-Hiroshima, 739-8526, Japan}
\affiliation{Interdisciplinary Theoretical and Mathematical Sciences Program (iTHEMS), RIKEN, Wako, Saitama 351-0198, Japan}
\author{Shigehiro Nagataki}
\affiliation{RIKEN Cluster for Pioneering Research (CPR), Astrophysical Big Bang Laboratory (ABBL), Wako, Saitama, 351-0198 Japan}
\affiliation{Interdisciplinary Theoretical and Mathematical Sciences Program (iTHEMS), RIKEN, Wako, Saitama 351-0198, Japan}
\author{Tomoya Hayata}
\affiliation{Departments of Physics, Keio University, 4-1-1 Hiyoshi, Kanagawa 223-8521, Japan}
\affiliation{Interdisciplinary Theoretical and Mathematical Sciences Program (iTHEMS), RIKEN, Wako, Saitama 351-0198, Japan}

\date{\today}
\begin{abstract}
We theoretically discuss the non-Hermitian superfluid phase transition in one-dimensional two-component Fermi gases near the {\it p}-wave Feshbach resonance accompanied by the two-body loss associated with the dipolar relaxation.
For the first time we point out that this system gives us an opportunity to explore the interplay among various non-trivial properties such as universal thermodynamics at divergent {\it p}-wave scattering length, topological phase transition at vanishing chemical potential, and non-Hermitian Bardeen-Cooper-Schrieffer(BCS) to Bose-Einstein condensate (BEC) transition, in a unified manner.
In the BCS phase, the loss-induced superfluid-normal transition occurs when the exceptional point appears in the effective non-Hermitian Hamiltonian.
In the BEC phase, the diffusive gapless mode can be regarded as a precursor of the instability of the superfluid state.
Moreover, we show that the superfluid state is fragile against the two-body loss near the topological phase transition point.
\end{abstract}
\maketitle

\section{Introduction}
The study of unconventional superconductors and superfluids has been an important subject in physics.
The elucidation of the pairing mechanism in these systems is a challenging but exciting problem in condensed matter physics.
One of their remarkable features is the anisotropy of the pairing order~\cite{RevModPhys.63.239}.
In particular, the {\it p}-wave pairing order has been regarded as a key for the application to fault-tolerant quantum computations~\cite{KITAEV20032}.
Moreover, it is widely believed that $^3P_2$ neutron superfluidity plays a crucial role in understanding neutron star physics such as the pulsar glitch and cooling processes by neutrino emissions~(for a review, see Ref. \cite{RevModPhys.75.607}). On the other hand, in these strongly-correlated systems, it is not easy to identify microscopic properties and engineer the pairing state due to their complicated structures, difficulties of observations, and disorders.

In this regard, ultracold atomic gases have attracted much attention as a quantum simulator of unconventional state of matter in strongly-correlated systems~\cite{RevModPhys.80.885}.
Indeed, ultracold atoms confined in magneto-optical traps are free from the unexpected disorder effects compared to condensed-matter systems.
Also, the interatomic interaction has experimentally been controllable by utilizing the Feshbach resonance~\cite{RevModPhys.82.1225}.
Very recently, the optical control of the {\it p}-wave Feshbach resonance has also been performed experimentally~\cite{PhysRevA.106.023322}.
Consequently, the evolution from the Bardeen-Cooper-Schrieffer (BCS) superfluid to molecular Bose-Einstein condensation (BEC) near the {\it p}-wave Feshbach resonance has been anticipated theoretically~\cite{PhysRevLett.94.050403,PhysRevLett.94.230403,PhysRevLett.96.040402,PhysRevLett.99.210402,PhysRevA.92.063638}.
However, realization of  the {\it p}-wave superfluid state in ultracold Fermi gases has been a long-standing issue.
The main obstacle is the strong atomic losses associated with the two-body dipolar relaxation as well as the three-body recombination~\cite{PhysRevLett.90.053201,PhysRevA.70.030702,PhysRevA.96.062704,PhysRevLett.101.100401,PhysRevLett.120.133401,PhysRevA.98.020702,PhysRevA.99.052704,welz2022anomalous,PhysRevA.104.043311}.  
Unfortunately, strongly interacting {\it p}-wave Fermi gases exhibit a short lifetime, and cannot reach the superfluid regime experimentally as of this moment because of the strong atomic losses.

\begin{figure}[t]
    \centering
    \includegraphics[width=0.7\linewidth]{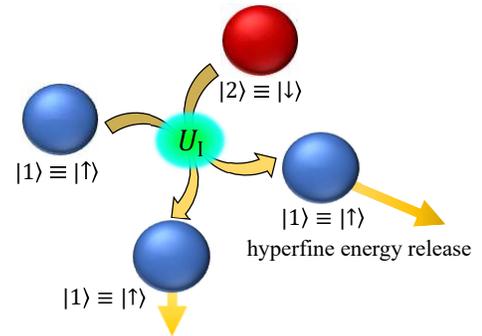}
    \caption{Two-body dipolar relaxation process leading the two-body loss near the $p$-wave Feshbach resonance~\cite{PhysRevA.96.062704}. The hyperfine transition from the excited state ($|2\rangle\equiv |\uparrow\rangle$) to the ground state ($|1\rangle\equiv |\downarrow\rangle$) leads to the hyperfine energy release involving the strong two-body loss of energetic atoms. Such a process can be characterized by the imaginary {\it p}-wave coupling constant $U_{\rm I}$~\cite{PhysRevA.95.032710}.}
    \label{fig:1}
\end{figure}

Towards the realization of {\it p}-wave superfluid Fermi gases, the suppression of two-body and three-body losses has been predicted in one dimension (1D)~\cite{PhysRevA.95.032710,PhysRevA.96.030701,PhysRevA.98.011603,PhysRevA.102.043319,PhysRevA.106.043310}.
Recently, the loss rates of 1D fermions near the {\it p}-wave resonance has been measured experimentally~\cite{PhysRevLett.125.263402,marcum2020suppression,jackson2022emergent}.
While it is still under debate how much the one-dimensional confinement reduces the {\it p}-wave atomic losses, 
it is important to investigate the stability of the {\it p}-wave superfluid state against the loss processes.
For such a purpose, the non-Hermitian description of open quantum systems may give a new insight on this problem~\cite{ashida2020non}.
Recently, the mean-field theory for the non-Hermitian {\it s}-wave pairing has been developed~\cite{PhysRevLett.123.123601,PhysRevA.103.013724,kanazawa2021non}.  
In addition, the imaginary part of the {\it p}-wave scattering volume, which is related to the two-body loss and the non-Hermitian two-body interaction, has been measured experimentally in {\it p}-wave two-component Fermi gases~\cite{PhysRevA.96.062704}.

In this work, we discuss the superfluid properties in 1D two-component Fermi gases near the {\it p}-wave Feshbach resonance accompanied by the two-body loss associated with the dipolar relaxation as shown in Fig.~\ref{fig:1}.  
Here, we consider the intercomponent {\it p}-wave interaction where the two-body loss process may be more important than that between identical fermions in spin-polarized systems. 
Even in the absence of the loss effect, the present system exhibits several unique properties such as the universal thermodynamics at the zero-range unitary limit~\cite{PhysRevA.104.023319} and the topological phase transition at vanishing chemical potential~\cite{PhysRevB.105.064508}.
We employ the non-Hermitian formalism involving the two-body losses characterized by the complex-valued two-body interaction.
Describing the framework of the non-Hermitian BCS-Leggett theory,
we discuss the stability of the superfluid state against the two-body loss near the {\it p}-wave resonance.

This paper is organized as follows. In Sec.~\ref{sec:2}, we present the non-Hermitian BCS-Leggett theory in 1D two-component Fermi gases with the intercomponent complex {\it p}-wave interaction. In Sec.~\ref{sec:3}, we investigate the exceptional point and the diffusive gapless mode, which play a crucial role for the non-Hermitian phase transition in this system.
In Sec.~\ref{sec:4}, we show the numerical results. We summarize this paper in Sec.~\ref{sec:5}. For convenience, we take $\hbar=k_{\rm B}=1$, and take the system length to be unity within the large system size limit.

\section{Non-Hermitian BCS-Leggett theory}
\label{sec:2}
We consider a 1D two-component Fermi gas with the intercomponent {\it p}-wave interaction 
and the inelastic two-body loss described by a quantum master equation $\frac{\partial \rho}{\partial t}=-i\left(H\rho-\rho H^\dag\right)$ without the quantum jump term~\cite{PhysRevLett.123.123601}, where $\rho$ is the density matrix.
The effective non-Hermitian Hamiltonian $H$ reads~\cite{PhysRevB.105.064508}
\begin{align}
\label{eq:H}
    &H=\sum_{k,\sigma}\xi_{k}c_{k,\sigma}^\dag c_{k,\sigma}\cr
    & \ +U\sum_{k,k',q}kk'
    c_{k+q/2,\uparrow}^\dag c_{-k+q/2,\downarrow}^\dag c_{-k'+q/2,\downarrow} c_{k'+q/2,\uparrow},
\end{align}
where
$\xi_{k}=\varepsilon_{k}-\mu\equiv\frac{k^2}{2m}-\mu$ is the kinetic energy of a fermion with a mass $m$ measured from the complex chemical potential $\mu=\mu_{\rm R}+i\mu_{\rm I}$.
$c_{k,\sigma}$ is an annihilation operator of a fermion with momentum $k$ and pseudospin $\sigma=\uparrow,\downarrow$.
The second term of Eq.~(\ref{eq:H}) is the {\it p}-wave interaction term describing the two-particle scattering with the center-of-mass momentum $q$ and the relative momenta $k$ and $k'$.
The complex coupling constant $U=U_{\rm R}-iU_{\rm I}$ is related to the complex {\it p}-wave scattering length as~\cite{PhysRevA.94.043636,PhysRevA.103.043307}
\begin{align}
    \frac{m}{2a}=\frac{1}{U}+\sum_{k}\frac{k^2}{2\varepsilon_k}.
\end{align}
Motivated by the experiment in 3D where the complex {\it p}-wave scattering volume has been measured~\cite{PhysRevA.96.062704}, we introduce the inverse {\it p}-wave scattering length~\cite{PhysRevA.95.032710}
\begin{align}
    \frac{1}{a}=\frac{1}{a_{\rm R}}+i\frac{1}{a_{\rm I}}.
\end{align}
We note that $a_{\rm R}\neq{\rm Re}[a]$, and $a_{\rm I}\neq{\rm Im}[a]$.
The explicit formula of the normalized coupling constant reads
\begin{align}
    \frac{m}{2a_{\rm R}}=\frac{U_{\rm R}}{|U|^2}+\frac{m\Lambda}{\pi},
    \quad
    \frac{m}{2a_{\rm I}}=\frac{U_{\rm I}}{|U|^2},
\end{align}
where $\Lambda$ is the momentum cutoff.
Here, we take $U_{\rm I}>0$ and then $1/a_{\rm I}>0$.
We briefly note that in the experiment for the 3D system~\cite{PhysRevA.96.062704}, the {\it p}-wave scattering volume is defined as $V^{-1}=V_{\rm R}^{-1}+iV_{\rm I}^{-1}$.
From the observed two-body loss rate,
the imaginary part is determined as $V_{\rm I}\simeq 4.8\times 10^{-21}$ m$^{3}$, while $V_{\rm R}$ is assumed to be tuned by applying the magnetic field. In this study, we take arbitrary combinations of $a_{\rm R}^{-1}$ and $a_{\rm I}^{-1}$ to understand the global phase diagram of the 1D counterpart.

To investigate the non-Hermitian many-body state,
we introduce a pair of the BCS-like ground states~\cite{PhysRevLett.123.123601}
\begin{align}
\label{eq:bcs1}
    |{\rm BCS}\rangle = \prod_{k}\left[u_k +v_{k}c_{k,\uparrow}^\dag c_{-k,\downarrow}^\dag
    \right]|0\rangle,
\end{align}
\begin{align}
\label{eq:bcs2}
    \langle\langle {\rm BCS}|=\langle 0|\prod_{k}
    \left[u_k +\bar{v}_{k} c_{-k,\downarrow}c_{k,\uparrow}
    \right],
\end{align}
where $u_k=\sqrt{\frac{1}{2}\left(1-\frac{\xi_k}{E_k}\right)}$, $v_k=-\sqrt{\frac{1}{2}\frac{\Delta(k)}{\bar{\Delta}(k)}\left(1-\frac{\xi_k}{E_k}\right)}$, and $\bar{v}_k=-\sqrt{\frac{1}{2}\frac{\bar{\Delta}(k)}{\Delta(k)}\left(1-\frac{\xi_k}{E_k}\right)}$~\cite{PhysRevA.103.013724,PhysRevLett.123.123601}.
Here $E_k=\sqrt{\xi_k^2+\Delta(k)\bar{\Delta}(k)}$ is the Bogoliubov quasiparticle dispersion with
the superfluid order parameters $\Delta(k)$ and $\bar{\Delta}(k)$ given by
\begin{align}
    \Delta(k)=-Uk\sum_{k'}k'\langle\langle c_{-k'\downarrow}c_{k'\uparrow}\rangle \equiv Dk,
\end{align}
\begin{align}
    \bar{\Delta}(k)=-Uk\sum_{k'}k'\langle\langle c_{k',\uparrow}^\dag c_{-k',\downarrow}^\dag \rangle \equiv \bar{D}k,
\end{align}
where $\langle\langle\cdots\rangle\equiv \langle\langle {\rm BCS}|\cdots|{\rm BCS}\rangle$ denotes the expectation value defined by Eqs.~(\ref{eq:bcs1}) and (\ref{eq:bcs2}). 
The resultant mean-field Hamiltonian reads
\begin{align}
\label{eq:hmf}
    H_{\rm MF}&=\sum_{k}\Psi_k^\dag
    \left(
    \begin{array}{cc}
        \xi_k &  -\Delta(k)\\
        -\bar{\Delta}(k) & -\xi_k
    \end{array}
    \right)
    \Psi_{k}-\frac{D\bar{D}}{U}+\sum_k\xi_k.
\end{align}
We note that the Hermitian counterpart (i.e., $U\in \mathbb{R}$) belongs to the Bogoliubov-de Gennes (BdG) class BDI~\cite{PhysRevB.78.195125}.
If the two-body loss is switched on, i.e., if $\mu$ and $D$ are complex, the non-Hermiticity breaks the time-reversal and chiral symmetries, and our BdG Hamiltonian belongs to the class D with a line gap and a sublattice symmetry $S_-$ (it is remnant of the chiral symmetry) according to the extended table in Ref.~\cite{PhysRevX.9.041015}. In one dimension, it is characterized by $\mathbb{Z}_2$ index~\cite{PhysRevX.9.041015}.
We obtain the ground-state energy as
\begin{align}
\label{eq:egs}
    E_{\rm GS}
    &=\sum_{k}\left[\xi_k-E_k+\frac{\bar{D}Dk^2}{2\varepsilon_k}\right]
    -\frac{m\bar{D}D}{2a}
\end{align}
where $E_k=\sqrt{\xi_k^2+D\bar{D}k^2}$.
We take $D\bar{D}=D_0^2$ and $H_{\rm MF}^\dag=H_{\rm MF}^*$ without loss of generality by taking an appropriate gauge transformation.
For convenience, we rewrite it as $D_0=D_{\rm R}+iD_{\rm I}$.
Accordingly, the complexified gap equation is given by $\frac{\partial E_{\rm GS}}{\partial D}=0$, that is,
\begin{align}
\label{eq:gapeq}
    \frac{m}{2a}+\sum_{k}k^2\left[\frac{1}{2E_k}-\frac{1}{2\varepsilon_k}\right]=0.
\end{align}
Also, the particle number density $N=-\frac{\partial E_{\rm GS}}{\partial \mu}$ reads
\begin{align}
\label{eq:numeq}
    N=\sum_{k}\left[1-\frac{\xi_k}{E_k}\right].
\end{align}
Practically, we solve Eqs.~\eqref{eq:gapeq} and \eqref{eq:numeq} with respect to four quantities $D_{\rm R}k_{\rm F}/E_{\rm F}$, $D_{\rm I}k_{\rm F}/E_{\rm F}$, $\mu_{\rm R}/E_{\rm F}$, and $\mu_{\rm I}/E_{\rm F}$
for a given real-valued number density $N\in \mathbb{R}$,
where $k_{\rm F}=\frac{\pi N}{2}$ and $E_{\rm F}=\frac{k_{\rm F}^2}{2m}$ are the Fermi momentum and energy, respectively.

Here, one may notice that Eqs.~(\ref{eq:gapeq}) and (\ref{eq:numeq}) are quite similar to those in the 3D {\it s}-wave pairing case~\cite{PhysRevA.103.013724}. Indeed, our framework shares several similar properties to the previous work.
In this regard, we restrict ourselves to the following situation
\begin{align}
\label{eq:dimui}
   D_{\rm R}, D_{\rm I}\geq 0,\quad \mu_{\rm I}\leq 0,
\end{align}
which are consistent with Ref.~\cite{PhysRevA.103.013724}.
However, $\mu_{\rm R}$ can be both positive and negative depending on the interaction parameters.

\section{Exceptional point and diffusive gapless mode}
\label{sec:3}
\begin{figure}[t]
    \centering
    \includegraphics[width=0.75\linewidth]{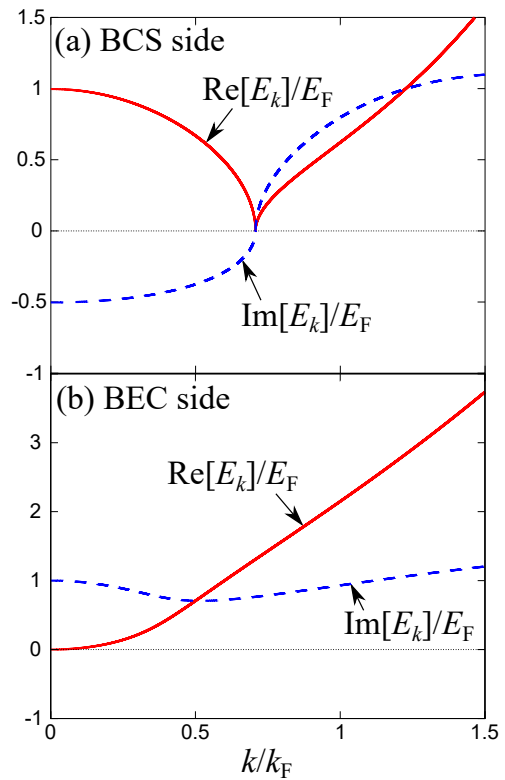}
    \caption{The complex dispersion relation $E_k$ (a) in the presence of the exceptional point in the BCS side and (b) at the appearance of the diffusive gapless mode in the BEC side. Here we use $\mu/E_{\rm F}=1-0.5i$ and $D_0k_{\rm F}/E_{\rm F}=\sqrt{2}(1+i)$ in the panel (a) and $\mu/E_{\rm F}=-i$ and $D_0k_{\rm F}/E_{\rm F}=2+0.5i$ in the panel (b). }
    \label{fig:2}
\end{figure}
Before showing the numerical calculation, we study the consequence of the exceptional point and the gapless excitation due to the two-body loss.
In the BCS side ($\mu_{\rm R}>0$), the superfluid solution disappears when the exceptional point appears in $H_{\rm MF}$ (i.e., two eigenvalues $\pm E_{k}$ coalesce~\cite{heiss2012physics}), leading to the divergence of $1/E_k$.
To see this,
we introduce
\begin{align}
    E_k^2=\xi_k^2 + D_0^2k^2 \equiv A_k+iB_k,  
\end{align}
where
\begin{align}
    A_k= \left(\varepsilon_k-\mu_{\rm R}\right)^2-\mu_{\rm I}^2 + (D_{\rm R}^2-D_{\rm I}^2)k^2,
\end{align}
\begin{align}
    B_k= 2D_{\rm R}D_{\rm I}k^2-2\mu_{\rm I}(\varepsilon_k-\mu_{\rm R}).
\end{align}
The exceptional point corresponds to $A_k=B_{k}=0$~\cite{PhysRevA.103.013724}.
In other words, the exceptional point can be found once $|k|$ satisfying $A_k=B_k=0$ exists.
In this regard, we introduce the singular momentum $k_{\rm S}$ satisfying $B_{k=k_{\rm S}}=0$ as 
\begin{align}
\label{eq:ks}
    k_{\rm S}=\bm{\pm}\sqrt{\frac{2\mu_{\rm R}\mu_{\rm I}}{\frac{\mu_{\rm I}}{m}-2D_{\rm R}D_{\rm I}}}.
\end{align}
The exceptional point appears when $A_{k=k_{\rm S}}=0$, that is,
\begin{align}
    -mD_{\rm I}D_{\rm R}+D_{\rm R}\sqrt{m^2D_{\rm I}^2+2m\mu_{\rm R}}=-\mu_{\rm I},
\end{align}
For convenience, we introduce the indicator of the exceptional point as
\begin{align}
\label{eq:z}
    z =D_{\rm R}\sqrt{m^2D_{\rm I}^2+2m\mu_{\rm R}}-mD_{\rm I}D_{\rm R}+\mu_{\rm I},
\end{align}
where the exceptional point appears at $z=0$.

Figure~\ref{fig:2}(a) shows the typical dispersion relation $E_k$ in the BCS side where the exceptional point appears at $k=k_{\rm S}$.
One can see that both ${\rm Re}[E_k]$ and ${\rm Im}[E_k]$ reach zero at $k=k_{\rm S}$, which leads to the disappearance of the superfluid solution in Eqs.~\eqref{eq:gapeq} and \eqref{eq:numeq}.

On the other hand, in the BEC side, the exceptional point never appears because $k_{\rm S}$ given by Eq.~\eqref{eq:ks} becomes pure imaginary when $\mu_{\rm R}<0$ with Eq.~(\ref{eq:dimui}).
However, $\mu_{\rm R}$ increases with increasing $(k_{\rm F}a_{\rm I})^{-1}$, and $\mu_{\rm R}=0$ may be achieved at a certain point, leading to the appearance of the singular momentum at $k_{\rm S}=0$.
In such a case, the negative $A_k\simeq -\mu_{\rm I}^2<0$ leads to the pure imaginary gapless dispersion relation $E_k=i\sqrt{|A_k|}$ at $k\simeq 0$.
Once such a diffusive gapless mode appears,
the calculation of the superfluid solution becomes numerically demanding due to the low-energy singularity.
Moreover, the expression of $E_{\rm GS}$ given by \eqref{eq:egs} is no longer valid because the gapless Bogoliubov excitation associated with the first term of Eq.~(\ref{eq:hmf}) plays a significant role at $\mu_{\rm R}>0$.
In this sense, the appearance of the diffusive gapless mode in the BEC side can be regarded as a signature of the phase transition where the superfluid solution disappears.
This feature is qualitatively consistent with the previous work in the 3D {\it s}-wave case~\cite{PhysRevA.103.013724} where the superfluid solution disappears in the lossy BEC regime when the two-body binding energy becomes pure imaginary.

Practically, it is numerically demanding to determine these transition points precisely both in the BCS and BEC sides because of the singular momentum integration in Eqs.~\eqref{eq:gapeq} and \eqref{eq:numeq}.
Therefore, in this work we fit the indicator $z$ in the BCS side and $\mu_{\rm R}$ in the BEC side with the linear function and extrapolate it to estimate the phase transition points.
We confirmed that this procedure does not change the results qualitatively.

\section{Numerical results}
\label{sec:4}
\begin{figure}[t]
    \centering
    \includegraphics[width=\linewidth]{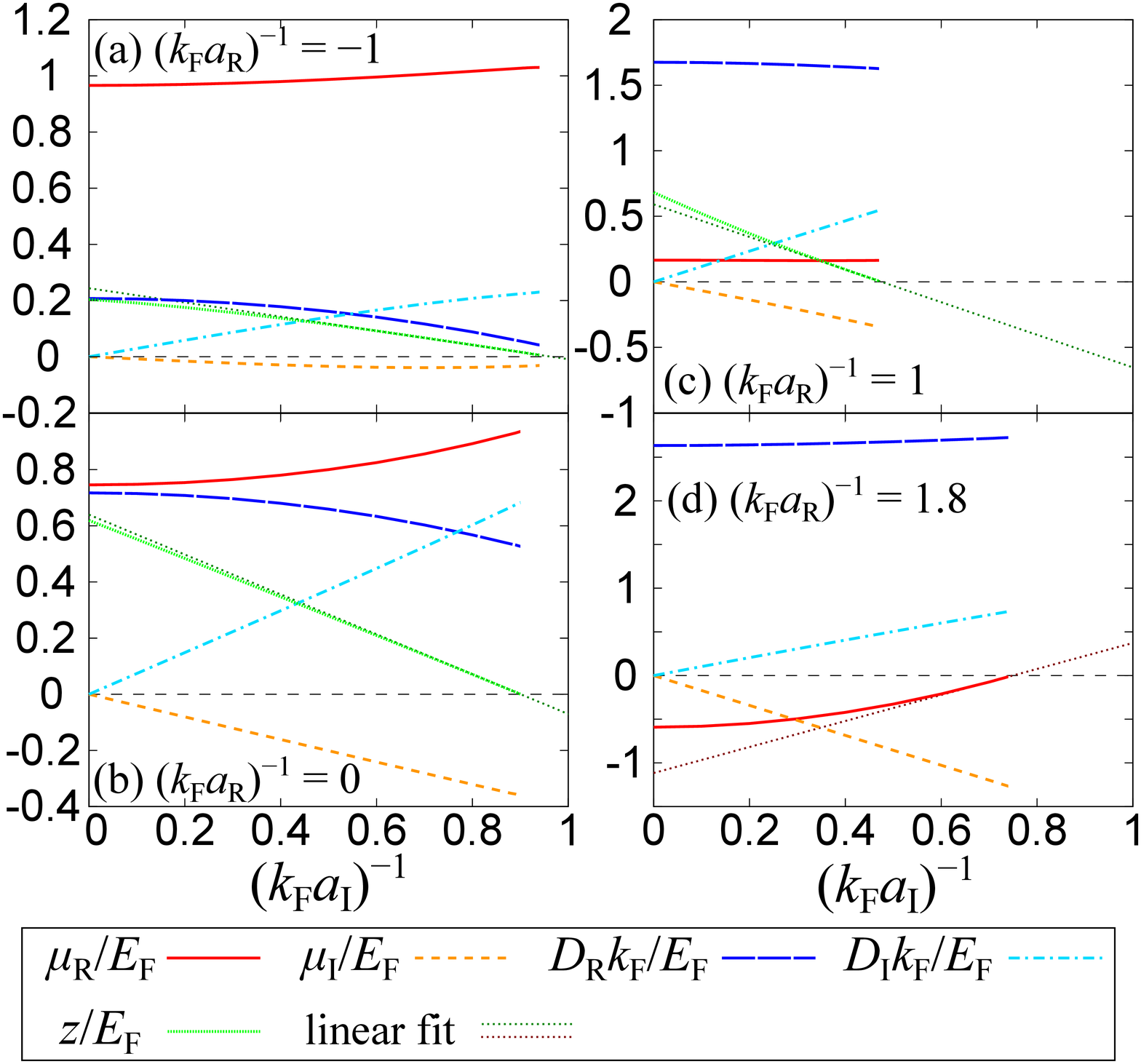}
    \caption{Calculated chemical potential $\mu=\mu_{\rm R}+i\mu_{\rm I}$ and superfluid order parameter $D_0=D_{\rm R}+iD_{\rm L}$ as a function of the two-body loss parameter $(k_{\rm F}a_{\rm I})^{-1}$ at (a)$(k_{\rm F}a_{\rm R})^{-1}=-1$, (b) $(k_{\rm F}a_{\rm R})^{-1}=0$, (c) $(k_{\rm F}a_{\rm R})^{-1}=1$, and (d) $(k_{\rm F}a_{\rm R})^{-1}=1.8$. While $z=0$ is regarded as the phase transition point in the panels (a), (b), and (c),
    $\mu_{\rm R}=0$ corresponds to the superfluid instability in the panel (d). To estimate these transition points, we used the linear fitting near the transition as plotted by the thin dotted lines.
    }
    \label{fig:3}
\end{figure}

Figure~\ref{fig:3} shows the numerical results of $\mu=\mu_{\rm R}+i\mu_{\rm I}$ and $D_0=D_{\rm R}+iD_{\rm I}$ at different interaction strength.
First, one can see general features of the present system, that is, the increase of $\mu_{\rm R}$ and $D_{\rm I}$, and the decrease of $\mu_{\rm I}$ when the two-body loss parameter $(k_{\rm F}a_{\rm I})^{-1}$ increases.
In particular, the increase of $\mu_{\rm R}$ may be regarded as a physical consequence where a larger $\mu_{\rm R}$ is required to realize a given $N$ under the two-body loss.
In the weak-coupling BCS regime, as shown in Fig.~\ref{fig:3}(a),
we find the numerical solution of $\mu$ and $D_0$ until the exceptional point given by $z=0$ appears.
Once the system is close to $z=0$, the numerical calculation of Eqs.~\eqref{eq:gapeq} and \eqref{eq:numeq} becomes severe and eventually, the solution disappears at $(k_{\rm F}a_{\rm I})^{-1}\simeq 0.9$, implying the instability of the superfluid phase.
Whenever $\mu_{\rm R}$ is positive, the exceptional point can be found at the {\it p}-wave unitary limit $(k_{\rm F}a_{\rm R})^{-1}=0$ shown in Fig.~\ref{fig:3}(b), as well as relatively strong-coupling regime shown in Fig.~\ref{fig:3}(c), apart from the quantitative difference of $\mu$ and $D_0$.
The exceptional point tends to appear at smaller $(k_{\rm F}a_{\rm I})^{-1}$ as  $(k_{\rm F}a_{\rm R})^{-1}$ gets larger in the BCS phase.
On the other hand, in the BEC phase with negative $\mu_{\rm R}$,
we do not encounter the exceptional point as explained in the previous section.
Instead, finding the superfluid solution becomes numerically demanding in the vicinity of the onset of the diffusive gapless mode, which would be a signal of the phase transition.
In this work, we identify this instability point at $\mu_{\rm R}=0$ by using the linear extrapolation (see Fig.~\ref{fig:3}(d) at $(k_{\rm F}a_{\rm R})^{-1}=1.8$).

\begin{figure}[t]
    \centering
    \includegraphics[width=0.9\linewidth]{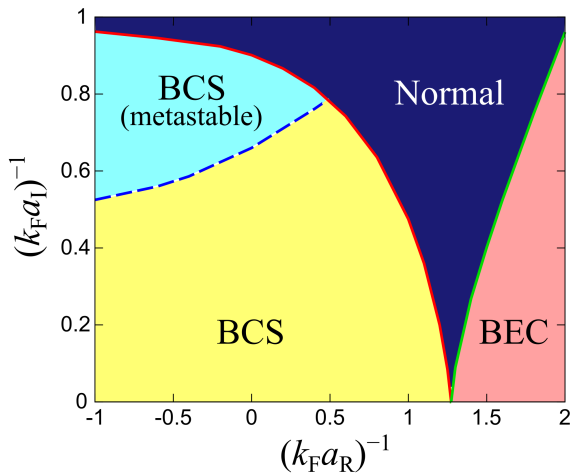}
    \caption{Phase diagram with respect to the inverse $p$-wave scattering length $(k_{\rm F}a)^{-1}=(k_{\rm F}a_{\rm R})^{-1}+i(k_{\rm F}a_{\rm I})^{-1}$. The phase boundary (red solid line) between the normal phase ``Normal" and the BCS phase ($\mu_{\rm R}>0$) ``BCS" (and metastable BCS) is given by the appearance of the exceptional point ($E_{k=k_{\rm S}}=0$). 
    The phase boundary (green solid line) between ``Normal" and the BEC phase ($\mu_{\rm R}< 0$) ``BEC" is given by the appearance of the diffusive gapless mode ($\mu_{\rm R}=0$).
    Also, the dashed line separating ``BCS" and ``BCS (metastable)" represents the first-order-like phase transition given by ${\rm Re}(F_{\rm S}-F_{\rm N})=0$, where $F_{\rm S}$ and $F_{\rm N}$ are the free energies in the superfluid phase and in the normal phase, respectively. 
    In this regard, the non-trivial superfluid solution is energetically stable in the ``BCS" and ``BEC" phases.
    }
    \label{fig:4}
\end{figure}

Figure~\ref{fig:4} shows the phase diagram in the plane of $(k_{\rm F}a_{\rm R})^{-1}$ and $(k_{\rm F}a_{\rm I})^{-1}$.
If $(k_{\rm F}a_{\rm I})^{-1}=0$ (i.e., the Hamiltonian is Hermitian),
the system exhibits a topological phase transition at $(k_{\rm F}a_{\rm R})^{-1}=4/\pi\simeq 1.27$~\cite{PhysRevB.105.064508},
where $\mu_{\rm R}$ changes the sign.
Thus, one may identify the BCS phase ($\mu_{\rm R}>0$) and the BEC phase ($\mu_{\rm R}<0$) by monitoring the sign of $\mu_{\rm R}$.
This fact is in stark contrast to the BCS-BEC crossover near the {\it s}-wave Feshbach resonance~\cite{ohashi2020bcs}.
According to Eq.~\eqref{eq:z}, this Hermitian topological phase transition at $(k_{\rm F}a_{\rm R})^{-1}=4/\pi$ can be regarded as an exceptional point (i.e., $z=0$) accompanied by the stable (real-valued) gapless mode $E_k= D_{0}|k|+O(|k|^3)$ (We note that this singular point in the Hermitian case is called a diabolic point in Ref.~\cite{ashida2020non}).
However, at $(k_{\rm F}a_{\rm I})^{-1}=0$, $E_k$ becomes zero only at $k=0$, and hence the divergence of $1/E_k$ do not cause the singularities in Eqs.~\eqref{eq:gapeq} and \eqref{eq:numeq}.

If $(k_{\rm F}a_{\rm I})^{-1}$ becomes finite, the gapless mode becomes diffusive and the exceptional point easily appears in the momentum space in such a regime.
In Fig.~\ref{fig:4}, the region beyond the upper bound given by $z=0$ in the BCS side and $\mu_{\rm R}=0$ in the BEC side is regarded as the normal phase because we cannot find the superfluid solution there.
In this sense, the region near the topological phase transition is fragile against the two-body loss.
In other words, the smooth crossover from the BCS phase to the BEC phase can be no longer expected in the present 1D {\it p}-wave system with the nonzero two-body loss in terms of the topology as well as the non-Hermicity.
On the one hand, at stronger coupling,
$\mu_{\rm R}$ becomes smaller due to the strong binding effect, leading to the stability against the two-body loss where the transition point $\mu_{\rm R}=0$ is retarded along the axis of $(k_{\rm F}a_{\rm I})^{-1}$.
The disappearance of the superfluid solution can be found even before the region where the real part of the two-body binding energy $E_{\rm b}$ changes its sign at $a_{\rm I}^{-1}=a_{\rm R}^{-1}$ (see Appendix~\ref{app:b}).
This result is in stark contrast to the 3D {\it s}-wave case where the superfluid solution disappears around ${\rm Re}[E_{\rm b}]=0$ in the BEC regime~\cite{PhysRevA.103.013724}.
Therefore, such a  disappearance of the superfluid solution
cannot be understood by the two-body physics but the many-body physics even in the strong-coupling regime.
We note that while the phase diagram in Fig.~\ref{fig:4} seems to be dramatically different from the previous work~\cite{PhysRevA.103.013724}, this is due to the definition of the complex scattering length, except for the presence of the topological phase transition at $(k_{\rm F}a)^{-1}=4/\pi$.
\begin{figure}
    \centering
    \includegraphics[width=7.5
cm]{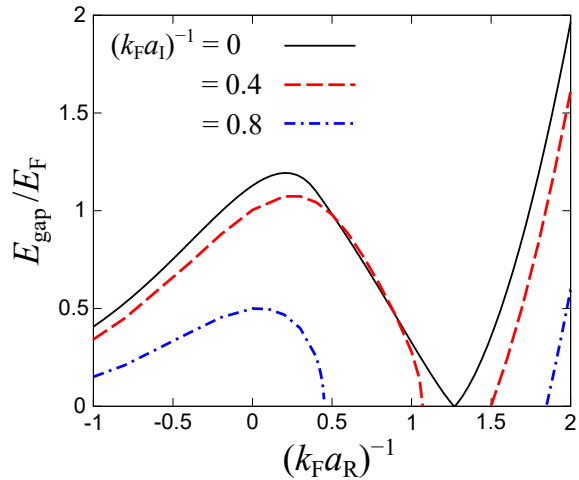}
    \caption{Excitation gap $E_{\rm gap}={\rm min}\left\{2{\rm Re}[E_k]\right\}$ at different $(k_{\rm F}a_{\rm I})^{-1}$. The gapless region with $E_{\rm gap}=0$ at nonzero $(k_{\rm F}a_{\rm I})^{-1}$ corresponds to the normal phase shown in Fig.~\ref{fig:4}. }
    \label{fig:5}
\end{figure}

On the other hand, at weaker coupling, $\mu_{\rm R}$ gradually approaches $E_{\rm F}$ and the spectrum exhibits the gapped excitation.
Figure~\ref{fig:5} shows the excitation gap defined by $E_{\rm gap}={\rm min}\left\{2{\rm Re}[E_k]\right\}$ at $(k_{\rm F}a_{\rm I})^{-1}=0$, $0.4$, and $0.8$.
In the Hermitian case [i.e., $(k_{\rm F}a_{\rm I})^{-1}=0$], it is given by $E_{\rm gap}=2D_0\sqrt{2m\mu-m^2D_0^2}$ at $\mu\geq mD_0^2$~\cite{PhysRevB.105.064508}.
Because of this excitation gap, the BCS phase is relatively stable against the two-body loss, but the superfluid solution disappears when the two-body loss is so strong that the exceptional point appears.
One can see that $E_{\rm gap}$ becomes smaller with increasing the two-body loss parameter. Eventually, the superfluid state is unstable at $E_{\rm gap}=0$. Indeed, this gapless points correspond to the appearance of the exceptional point in the BCS phase and the diffusive gapless mode in the BEC phase, respectively.
While we discuss only the specific system,
our result indicates that the gapless Fermi superfluid would be fragile against the two-body loss.

In this way, the zero-range {\it p}-wave unitary limit is protected by the excitation gap
and one may expect the realization of a {\it p}-wave unitary Fermi gas~\cite{PhysRevA.104.023319} even under the relatively small two-body loss.
Our mean-field result indicates that the corresponding Bertsch parameter defined via $\mu_{\rm R}/E_{\rm F}$ is enlarged by the two-body loss as 
\begin{align}
\xi_{\rm B}\simeq 0.75+0.22(k_{\rm F}a_{\rm I})^{-2}.
\end{align}
We note that the increasing behavior of $\mu_{\rm R}/E_{\rm F}$ with $(k_{\rm F}a_{\rm I})^{-1}$ is common at the different interaction strengths as shown in Fig.~\ref{fig:3}.
While the transdimensional equivalence of the Bertsch parameters between 1D {\it p}-wave and 3D {\it s}-wave unitary Fermi gases has been conjectured in Ref.~\cite{PhysRevA.104.023319}, our result $\xi_{\rm B}\simeq 0.75$ at $(k_{\rm F}a_{\rm I})^{-1}=0$ is larger than the experimental value $\xi_{\rm B}\simeq0.37$~\cite{ku2012revealing,PhysRevX.7.041004} and moreover the mean-field result $\xi_{\rm B}\simeq0.595$~\cite{ohashi2020bcs} in 3D unitary gases. To examine the transdimensional equivalence in more detail, it is important to address the beyond-mean-field effect such as self-energy shift associated with the many-body $T$-matrix~\cite{PhysRevA.85.012701,PhysRevA.95.043625}, which is beyond the scope of this paper.

Finally, under the strong two-body loss, the superfluid state may become metastable even before reaching the exceptional point~\cite{PhysRevA.103.013724}.
To this end, we introduce the free energy $F_{\rm S}=E_{\rm GS}+\mu N$.
In the normal phase ($D_0=0$),
we obtain the free energy $F_{\rm N}=\frac{k_{\rm F}^3}{3\pi m}$.
If ${\rm Re}[F_{\rm S}-F_{\rm N}]>0$, the superfluid state is regarded as a metastable state.
In this regard, we define the first-order-like transition point as ${\rm Re}[F_{\rm S}-F_{\rm N}]=0$.
In Fig,~\ref{fig:4},
we also plot this transition point, beyond which the metastable BCS state can be found. 
Although the definition of the scattering parameter is different from the previous work on the 3D {\it s}-wave case~\cite{PhysRevA.103.013724},
the similar transition can be found therein.

\section{Summary}
\label{sec:5}
In this work, we have theoretically investigated the superfluid phase transition of a 1D non-Hermitian {\it p}-wave unitary Fermi gas.
We developed the non-Hermitian BCS-Leggett theory for the present system and identified when and how the superfluid solution disappears due to the two-body loss term, by solving the gap equation and number equation numerically.

First, we pointed out that in the absence of the two-body loss, the BCS and BEC phases are distinguished by the sign of the real part of the chemical potential $\mu_{\rm R}$.
In the presence of the two-body loss,
the superfluid solution can be found up to the occurrence of the exceptional point $E_k=0$ in the BCS phase and the diffusive gapless mode with $\mu_{\rm R}\rightarrow -0$ and $\mu_{\rm I}<0$ in the BEC phase.
Beyond these points, the solution of the non-Hermitian superfluid disappears and thus we concluded that the superfluid state becomes unstable due to the two-body loss if a gapless mode exists.
In this sense, we found that the superfluid is particularly fragile near the topological phase transition at $(k_{\rm F}a_{\rm R})^{-1}=4/\pi$ against the two-body loss.
Our non-Hermitian framework suggests that it would be better to address the superfluid state far away from the topological phase transition point in future experiments.
In the BCS regime, before reaching the exceptional point,
the system may undergo the first-order-like phase transition from the BCS phase to the normal phase with increasing the two-body loss.
We showed such a transition point by monitoring the real part of the free energy.

Regarding the future perspective,
while we show the stability of the superfluid state in the presence of the two-body loss within the non-Hermitian BCS-Leggett formalism, it would be important to include effects of the jump term in the Lindblad equation and study dynamical evolution of the system.
While we worked in the 1D system, a similar non-Hermitian quantum phase transition can be studied in the 3D system near the {\it p}-wave resonance, where the gapless mode plays a crucial role~\cite{PhysRevLett.94.230403}.
Furthermore, the three-body loss term should be included in the future study.
It would also be interesting to study pairing fluctuation effects in the non-Hermitian systems.
The phase transition associated with the exceptional point may be probed via the optical spin conductivity measurement~\cite{PhysRevB.105.064508,PhysRevResearch.4.043014}.

\acknowledgements
The authors thank S. Uchino for useful discussion in the early stage of this study.
H.~T. was supported by the JSPS Grants-in-Aid for Scientific Research under Grant Nos.~18H05406, ~22K13981, ~22H01158.
Y.~S. and S.~N. are supported by the RIKEN Pioneering Project: Evolution of Matter in the Universe (r-EMU). A.~D. was supported by JSPS Research Fellowship for Young Scientists (22J10448).
T.~H. was supported by JSPS KAKENHI Grant Numbers~21H01007, and 21H01084.

\appendix

\section{two-body bound state}
\label{app:b}
The two-body $T$-matrix reads
\begin{align}
    T(k,k';\omega)=Ukk'+\sum_{p}\frac{Ukp}{\omega_+-p^2/m} T(p,k';\omega).
\end{align}
Assuming the separability $T(k,k;\omega)=t(\omega)kk'$,
we obtain
\begin{align}
    t(\omega)
    &=U\left[1-U\sum_{p}\frac{mp^2}{m\omega_+-p^2}\right]^{-1}.
\end{align}
The bound state can be found as the pole of the $T$-matrix $t^{-1}(\omega=-E_{\rm b})=0$ given by
\begin{align}
    \frac{1}{U}+\sum_{p}\frac{mp^2}{p^2+mE_{\rm b}}=0.
\end{align}
Therefore, we obtain
$\frac{m}{2a}-\frac{m}{2}\sqrt{mE_{\rm b}}=0$, leading to $E_{\rm b}=\frac{1}{ma^2}.$
More explicitly,
\begin{align}
    E_{\rm b}
    &=\frac{1}{ma_{\rm R}^2}-\frac{1}{ma_{\rm I}^2}+\frac{2}{ma_{\rm R}a_{\rm I}}i.
\end{align}
In this regard, the real part of the binding energy ${\rm Re}[E_{\rm b}]$ becomes negative at $a_{\rm R}^{-1}<a_{\rm I}^{-1}$. 


\bibliographystyle{apsrev4-2}
\bibliography{reference.bib}

\end{document}